\documentclass[useAMS,usenatbib]{mn2e}

\usepackage{epsfig}
\usepackage{amsmath}
\usepackage{natbib}
\usepackage{times}
\usepackage{amsfonts}
\usepackage{aas_macros}
\usepackage{url}

\title[MACS High-Redshift Galaxies]{A Systematic Search for Lensed High-Redshift Galaxies in HST Images of MACS Clusters}
\author[Repp et al.]{A. Repp$^{1}$, H. Ebeling$^1$, and J. Richard$^2$\\
$^{1}$Institute for Astronomy, 2680 Woodlawn Dr., Honolulu, Hawaii 96822, USA\\
$^{2}$CRAL, Observatoire de Lyon, Universit\'{e} Lyon 1, 9 Avenue Ch. Andr\'{e}, 69561 Saint Genis Laval Cedex, France}

\begin{document}

\date{\today}

\pagerange{\pageref{firstpage}--\pageref{lastpage}} \pubyear{0000}

\maketitle

\label{firstpage}

\begin{abstract}
We present the results of a 135-arcmin$^2$ search for high-redshift galaxies lensed by 29 clusters from the MAssive Cluster and extended MAssive Cluster
Surveys (MACS and eMACS). We use relatively shallow images obtained with the Hubble Space Telescope in four passbands, namely,
F606W, F814W, F110W, and F140W. We identify 130 F814W dropouts as candidates for galaxies at  $z \ga 6$. In order to fit
the available broad-band photometry to galaxy spectral energy distribution (SED) templates, we develop a prior for the level
of dust extinction at various redshifts. We also investigate the systematic biases incurred by the
use of SED-fit software.
The fits we obtain yield an estimate of 20 Lyman-break galaxies with
photometric redshifts from $z \sim 7$ to 9. In addition, our survey has identified over 100 candidates with
a significant probability of being lower-redshift ($z \sim 2$) interlopers. We conclude that even as few as
four broad-band filters -- when combined with fitting the SEDs -- are capable of isolating promising objects. Such surveys thus allow one both to probe the bright end ($M_{1500} \la -19$) of the 
high-redshift UV luminosity function and to identify candidate massive evolved galaxies at lower redshifts.
\end{abstract}

\section{Introduction}
Reionization and galaxy formation are key events in cosmic history; the former
depends upon, and the latter is responsible for, the characteristics of high-redshift galaxies. Therefore
the study of such galaxies is a major component of the ongoing intensive investigation into the early
epochs of the universe.

The most credible redshift determinations arise from spectroscopic analysis. Although the catalog of
spectroscopically confirmed high-redshift galaxies contines to expand (e.g., \citealt{Richard2011,
 Vanzella2011, Bradac2012, Bradley2012a, Ono2012, Finkelstein2013}, \citealt{Oesch2015, Zitrin2015}), the time required to obtain spectra
of faint objects limits the scope of spectroscopic surveys.
Hence the most fruitful method for expanding the catalog of high-redshift galaxies is
the dropout technique \citep{Steidel1995, Steidel1996}, which relies on multiple-passband
photometry. Breaks in an object's spectrum -- in particular, both the Lyman break (at restframe 912 \AA) and the 4000-\AA\ break -- can cause it to `drop out' of passbands blueward of
the break due to absorption of its radiation by neutral gas. Comparison of the observed and rest-frame
wavelengths of the break immediately yields a crude (`photometric') redshift estimate.
One can subsequently improve this estimate by fitting galaxy template spectral energy distributions (SEDs)
to the observed multi-band photometry.

Multiple researchers have applied this technique to identify both intermediate-redshift
\citep{Steidel1999, Ellis2001, Giavalisco2004, Ouchi2004a,
Stark2009, Oesch2010} and high-redshift galaxies \citep{Beckwith2006, Bouwens2006, Bouwens2010a,
Bouwens2011,Lorenzoni2013, Oesch2013, Schenker2013, Ellis2013}. The widest of these
surveys have covered up to 1.65 deg$^2$ (e.g., \citealt{Bowler2014}) 

To push this technique to fainter magnitudes, other studies have combined it with the power of gravitational lensing
\citep{Ellis2001, Richard2006, Richard2008, Bradley2013, Atek2014, Zheng2014}; it was thus
that \cite{Coe2013} identified a galaxy with photometric redshift $z \sim 11$ (see also
\citealt{Coe2015}).
Lensed surveys tend to cover a smaller solid angle than field surveys because of
their dependence on high-mass foreground galaxy clusters. One of the most
extensive such projects is the Cluster
Lensing And Supernova survey with Hubble (CLASH---\citealt{Postman2012}), which imaged
25 clusters in 16 filters, with integration times in each filter ranging from 1975 to 4920 s. (See for instance
\citealp{Bradley2013, Zitrin2013, Bouwens2014}.)
Most lensed
surveys apply more time to smaller solid angles; for instance, the Hubble Frontier
Fields\footnote{\url{http://www.stsci.edu/hst/campaigns/frontier-fields}}
program devotes 140 orbits to each of six massive clusters. Still ongoing, this deep-imaging program
has already detected a substantial number of high-redshift galaxies (see \citealt{Atek2014,Atek2015,
Zheng2014,McLeod2015,Ishigaki2015}).

Despite these successes, photometric redshifts always contain
an element of uncertainty, given the possibility of low-redshift objects mimicking the colours
of high-redshift galaxies \citep{Mobasher2005, Schaerer2007, Schenker2012}.
These low-redshift interlopers may be
red stars or galaxies with high equivalent-width emission lines
\citep{Atek2011}. Thus in searching for high-redshift dropouts, it is important to analyse passbands blueward of the dropout
band in order to detect the flux enhancement due to strong emission lines (like H$\alpha$ and [OIII])
which would indicate a lower redshift.

To summarize, one can increase the yield of photometric redshift surveys by employment of larger
sample sizes, by utilization of gravitational lensing from massive galaxy clusters, and by a judicious choice of passbands.
The most massive clusters known to date at $z > 0.3$ are those identified
by the MAssive Cluster Survey (MACS) \citep{Ebeling2001, Ebeling2007,
Ebeling2010, Mann2012}, which
systematically catalogued the most X-ray luminous -- and hence the most massive -- galaxy
clusters. In this work we employ the lensing power of 29 such clusters
not studied by CLASH, thus conducting one of the broadest lensed dropout
searches to date. Table~\ref{tab:exposures} compares this survey to CLASH; despite
the longer integration times and the greater number of passbands in CLASH, the
similarity in solid angle coverage bodes well for the identification
high-redshift candidates by our survey.

Throughout this paper we assume a standard concordance
cosmology with $H_0=70$ km s$^{-1}$ Mpc$^{-1}$, $\Omega_m=0.3$, and
$\Omega_\Lambda=0.7$. We express all magnitudes in the AB system \citep{ABref}.

\section{Data}
\label{sec:data}
This survey analyses Hubble Space Telescope (HST) images of the 28 MACS clusters and one eMACS (extended MACS) cluster \citep{Ebeling2013}
listed in Table~\ref{tab:clusters}. This sample comprises all MACS clusters that were not part of CLASH
and for which HST images exist in all of the following
passbands: F606W and F814W on the Advanced Camera for Surveys (ACS); and F110W and F140W
on the Wide Field Camera 3 (WFC3). All images were obtained through HST SNAP-shot
programs\footnote{GO-10491, GO-10875, GO-12166, GO-12884: PI H.Ebeling}. 
The exposure times (identical within each passband across all clusters) appear in Table~\ref{tab:exposures}.

\begin{table}
  \centering
  \begin{minipage}{8.5cm}
  \caption{Exposure times}
  \label{tab:exposures}
  \begin{tabular}{lcccccc} \hline
            & \multicolumn{4}{c} \dotfill\mbox{Integration Times}\dotfill  & Fil-  &  Clus- \\ 
            &  F606W   &  F814W   &  F110W   &  F140W   & ters  			& ters\\ \hline
 This work      &         1200 s    &  1440 s    &    706 s   &   706 s    & 4 		&   29\\
 CLASH$^a$   &         1975 s    & 4103 s    &  2415 s    &  2342 s    & 16              &   25\\
\hline
 \end{tabular}
$^a$ CLASH integration times are averages from table~5 of \citet{Postman2012}.
Actual exposure times vary from cluster to cluster, depending on the quality of previous observations.
\end{minipage}
\end{table}

\begin{table}

  \begin{minipage}{8.5cm}
  \centering
  \caption{Clusters surveyed, with Milky Way hydrogen column density (in units of $10^{20}$ cm$^{-2}$)
   for estimating extinction}
  \label{tab:clusters}
  \begin{tabular}{lclc} \hline
 Cluster			&	$n_H$	&	 Cluster	&	$n_H$\\\hline
eMACSJ1057.5+5759	&	0.56	&MACSJ1354.6+7715	&	2.86\\	      
MACSJ0140.0$-$0555	&	2.85	&MACSJ1621.4+3810	&	1.12\\	
MACSJ0152.5$-$2852	&	1.51	&MACSJ1652.3+5534	&	2.33\\	
MACSJ0257.6$-$2209	&	2.18	&MACSJ1731.6+2252	&	6.29\\	
MACSJ0451.9+0006	&	7.23	&MACSJ1738.1+6006	&	3.74\\	
MACSJ0712.3+5931	&	5.43	&MACSJ1752.0+4440	&	3.06\\	
MACSJ0916.1$-$0023	&	3.25	&MACSJ2050.7+0123	&	7.50\\	 
MACSJ0947.2+7623	&	2.22	&MACSJ2051.1+0215	&	8.14\\	
MACSJ1115.2+5320	&	0.90	&MACSJ2135.2$-$0102	&	4.27\\	
MACSJ1124.5+4351	&	2.04	&SMACSJ0234.7$-$5831	&	3.64\\	
MACSJ1133.2+5008	&	1.44	&SMACSJ0549.3$-$6205	&	4.50\\	   
MACSJ1142.4+5831	&	1.77	&SMACSJ0600.2$-$4353	&	6.24\\	
MACSJ1226.8+2153C	&	1.87	&SMACSJ2031.8$-$4036	&	3.91\\	
MACSJ1236.9+6311	&	1.68	&SMACSJ2131.1$-$4019	&	3.00\\	
MACSJ1319.9+7003	&	1.47	&\\
 \hline
\end{tabular}
\end{minipage}
\end{table}

The WFC3 field of view (4.65 arcmin$^2$), being smaller than that of the ACS, determines
the 135-arcmin$^2$ solid angle of this survey.

\section{Analysis}
\label{sec:analysis}
One result of the opportunistic nature of the SNAP-shot program is that the four images of each
cluster are taken at random times dictated by scheduling requirements. As a result the images of 
a given cluster are not, in general, aligned. In addition, the ACS plate scale is significantly smaller than that
of WFC3. In order to facilitate comparison between images in different passbands, we redrizzled all images
(using \texttt{DrizzlePac}\footnote{\url{http://drizzlepac.stsci.edu}})
to the pixel scale and reference frame defined by the F140W image for the relevant cluster. 

The presence of low-sensitivity
regions (`blobs' -- see \citealt{WFC3instrHB}) can complicate the analysis of WFC3 images.
These artefacts are the result of differential reflectivity of the Channel Select Mechanism Mirror.
In several instances (discussed in Section~\ref{sec:results}) one of these blobs (or another defect) in the F110W image
coincides with both a detection in the F140W channel and dropout behaviour in F814W and F606W
(see for instance the final row of Fig.~\ref{fig:galinfo2}). In addition, ambient light
had contaminated the majority of the F110W image for cluster MACSJ0712.3+5931 (see for
instance the fourth and fifth rows of Fig.~\ref{fig:galinfo1}). In both of these situations,
only three passbands of usable data are available.

We then stacked the F140W and F110W images and ran \texttt{SExtractor} \citep{SExt} in dual-image mode, using the
stacked image for detection. We employed a 12-pixel
rectangular annulus for background determination and thus obtained a catalog of objects with isophotal
magnitudes in each passband. Our inital catalog of SExtractor results comprise 37,809 records, the flux errors of which
we corrected for correlated noise according to the prescription of \citet{Casertano}. Since the
WFC3 images determine our segmentation, and since the WFC3 point-spread function (PSF) is
significantly wider than that of the ACS, we do not perform any PSF-matching. We want to capture as
much flux as possible from the ACS images in order to insure that the objects we consider are truly
dropping out in the ACS bands. By retaining the tighter ACS PSF we allow for more accurate detection of this dropout behavior.

\begin{figure}
\leavevmode
\epsfxsize=8.5cm
\epsfbox{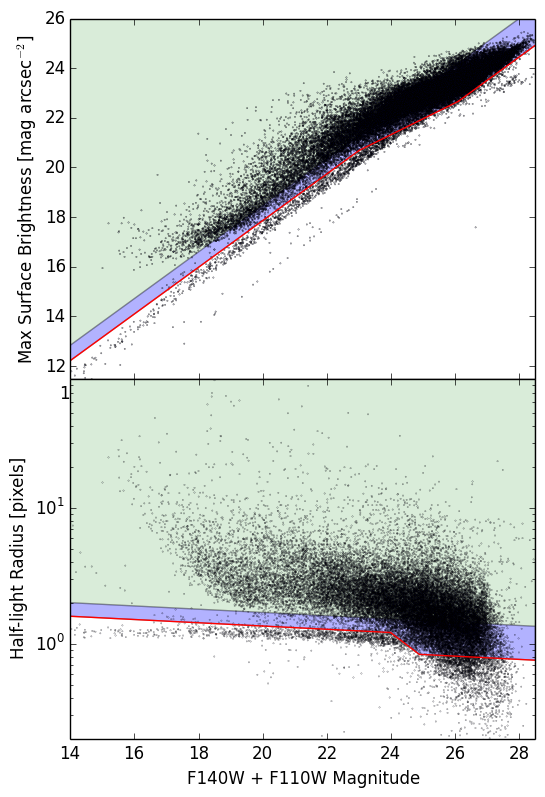}   
\caption{We use both morphology and spectral energy distribution to discriminate between stars and galaxies; 
these plots show our morphological criteria. In both panels the line corresponding
to point sources (stars) is apparent, running diagonally in the top panel and (roughly) horizontally in the bottom panel.
Any object below this line is more compact than a point source and thus must be an artefact. Hence we
discard anything in the unshaded portion of either plot. Since the star line and the
galaxy cloud begin to interpenetrate at higher magnitudes and since HST may not resolve compact faint
high-redshift galaxies, we place this cutoff $1\sigma$ above the star line at low magnitudes and $2\sigma$
below the star line at high magnitudes. Any object in the light green (upper) region of both plots ($3\sigma$
above the star line) we accept as a galaxy without further examination. The remaining objects (in the middle, blue-shaded
region) we subject to the spectral criterion outlined in Section~\ref{sec:spectest}.}
\label{fig:stargal}
\end{figure}

The next task is to discriminate between galaxies, stars, and artefacts. To do so, we consider both the 
objects' morphology and their colours; we describe the morphological criterion here and the spectral
criterion in Section~\ref{sec:spectest}. As Fig.~\ref{fig:stargal} shows, point sources (stars) occupy
a well-defined region (`star line') in both magnitude/surface brightness space and magnitude/half-light
radius space. Any object which lies more than $3\sigma$ above this line in both plots (i.e., in the light green regions
of both panels of Fig.~\ref{fig:stargal}) we accept as a galaxy without further examination.

Ideally we could now reject anything on this line as a star and anything below this line as an artefact. However,
at higher magnitudes the galaxies begin to bleed into the star line; in addition, high-redshift galaxies can
remain unresolved by WFC3 \citep{Oesch2010}. To account for these facts we place the rejection limit $1\sigma$ above the
star line at low magnitudes but $2\sigma$ below the star line at high magnitudes; anything that lies below this limit in
either plot (i.e., in the unshaded region of either panel of Fig.~\ref{fig:stargal}) we reject as being either a star or
an artefact.

The remaining objects lie in the blue region of the figure; we provisionally admit these objects
into the next phase of our analysis but will use the spectral criterion of Section~\ref{sec:spectest}
to eliminate M-stars and brown dwarfs.

At this point we also exclude detections in the noisy portions of the WFC3 field of view.\footnote{We exclude detections within fifteen pixels
of the edge of the frame as well as in the defect near the bottom of the detector (dubbed the
`death star' in \citealt{WFC3instrHB}), in addition to overexposed regions, defined
as any region with a surface brightness of less than 15 magnitudes arcsec$^{-2}$.}

We identify F814W dropouts by requiring at least a 5$\sigma$ detection in either WFC3 band
(F140W or F110W) and less than 2$\sigma$ signal-to-noise ratio in both ACS bands (F814W and F606W).
We visually inspected the dropouts, eliminating diffraction spikes and areas in which a nearby bright source
had corrupted the SExtractor results; we also eliminated `detections' that appeared to be 
serendipitously grouped noise. 

In addition, we checked the SExtractor segmentation map for these sources to ensure that SExtractor
properly discriminated between the objects themselves and neighboring sources. For cases in which
it did not, we set an appropriate aperture for each object and reperformed the photometry in those
apertures. In some cases the aperture photometry resulted in at least a 2$\sigma$
detection in the dropout band, causing us to eliminate these sources from consideration. Our final
$I_{814}$-dropout catalog consists of 130 sources.

Finally, to account for foreground (Milky Way) extinction, we first convert the column densities $n_H$
from Table~\ref{tab:clusters} to values of $A_V$ using the prescription of \citet{Guver2009} and
then apply the \citet{Cardelli1989} law to calculate the extinction in each band.

\section{SED fitting}
\label{sec:SEDfit}

\subsection{Obtaining Redshift Probability Distributions}
Both the Lyman break and the 4000-\AA\ break
can cause dropout behaviour. Thus, a
decrease in F814W flux accompanied by a strong detection in F110W could reflect the 4000-\AA\ break
redshifted to $1.5 \la z \la 2$ or the Lyman break redshifted to $7 \la z \la 10$.
The lower-redshift objects are likely to be massive, passively-evolving galaxies. These galaxies
(at such redshifts) typically are quiescent, extremely compact, and already quite old, with mass
densities at least an order of magnitude greater than those of local elliptical galaxies
\citep{Toft2012, Toft2014}.
Some of these objects appear to be the cores about which
the most massive of today's galaxies were built \citep{vanDokkum2014}. Since evolved $z\sim2$
galaxies can serve as observational proxies for similar objects at higher redshifts, these
interlopers are themselves promising candidates for future study.

However, our primary interest for this work is high-redshift galaxies; thus, to exclude low-redshift
objects we fit galaxy template spectral energy distributions (SEDs) to the observed fluxes
to obtain photometric redshifts for our candidates.

For this purpose we use \texttt{BPZ}\footnote{\url{http://www.stsci.edu/~dcoe/BPZ/}}
(Bayesian Photometric Redshift: \citealt{BPZ1, BPZ2, BPZ3}), which matches
an object's SED to known galactic spectral types and produces a
probability distribution for the object's redshift.

\begin{figure}
    \leavevmode\epsfxsize=10cm\epsfbox{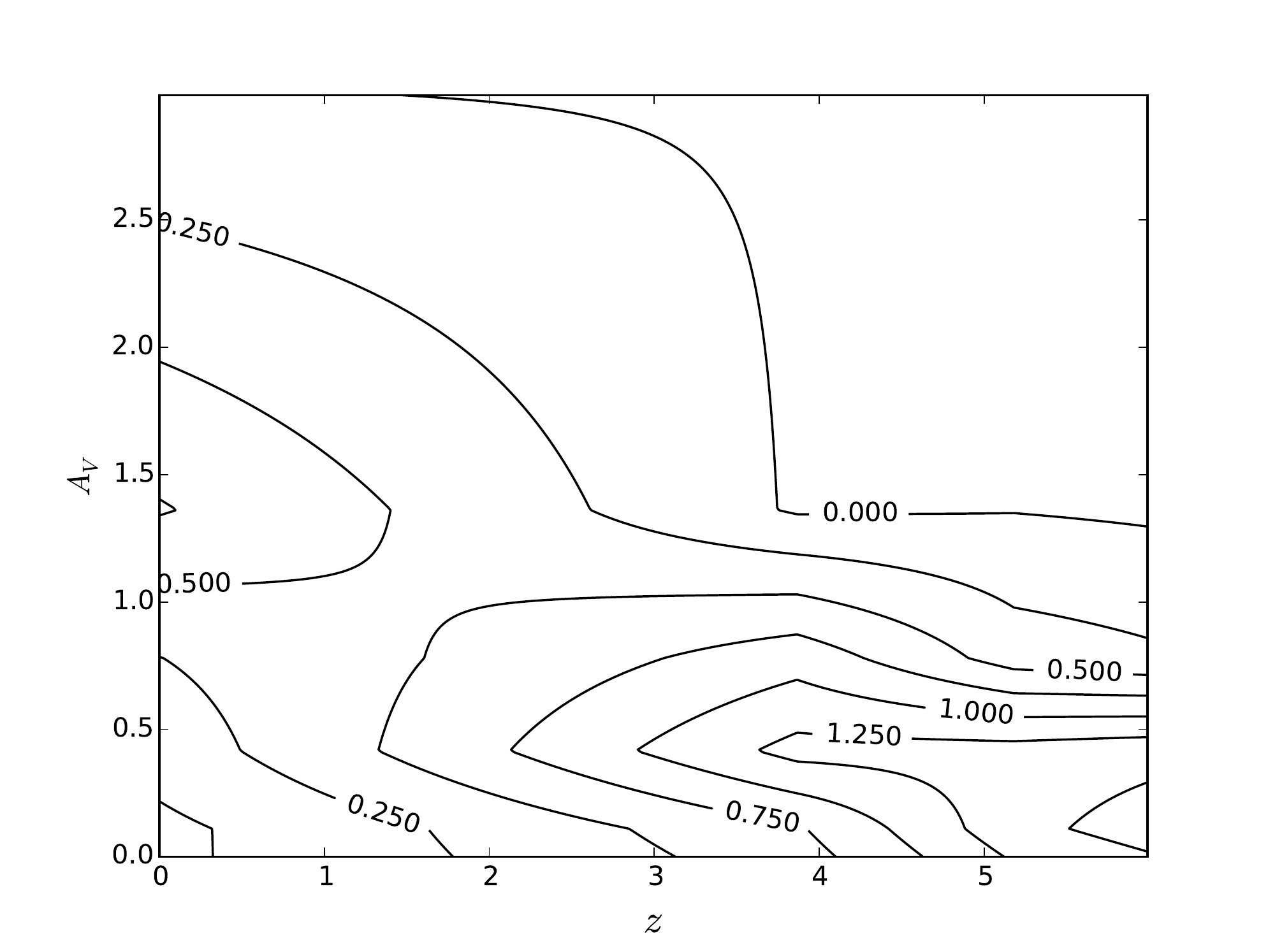}
    \caption{Contours showing prior probability distribution for intrinsic extinction $A_V$ at a given
redshift $z$, derived from data in \citet{Bouwens2009} and \citet{Magdis2010}.}
\label{fig:dustprior}
\end{figure}

However, BPZ's default templates do not allow one to include extinction as a separate
parameter, although they do empirically reproduce the observed photometry of the wide variety
of galaxies observed in large-scale surveys (see, for example, \citealt{Rafelski2015}). We deemed
it advisable to explicitly include a prior for intrinsic exinction, given the incidence of galaxies with
non-neglible dust attenuation (e.g., \citealp{Boone2011,Dey1999}). We thus
obtain dust-extinguished templates by 
applying the \citet{Calzetti2000} extinction law to each default template
in increments of $\Delta A_V = 0.5$ from $A_V = 0$ to 3.
Not all extinction values occur with equal probability, however, and the probability of extinction
evolves with redshift. We therefore require a Bayesian prior for the likelihood of
a given extinction $A_V$ at a given redshift. To obtain such a prior we employ the data from
fig.~6 of \citet{Bouwens2009}, who use observed UV continuum slopes
$\beta$ for a sample of Lyman break galaxies to derive estimates of $E(B-V)$
for redshifts from 2.5 to 6.
\citet{Bouwens2009} also
estimate the effective selection volume at each redshift, allowing calculation of
the percentage of galaxies with a given intrinsic extinction (as a function of redshift).
We extend these probabilities to the local universe using \citet{Magdis2010}'s
estimate that galaxies are 8 to 10 times less obscured at $z \sim 2$ than they are now; and
we extend them to higher redshifts by assuming the same probabilities obtained at
$z \sim 6$. We then
interpolate, smooth, and normalize the resulting distribution to obtain the prior shown
in Fig.~\ref{fig:dustprior}. One can obtain an analytic estimate for this distribution
(Repp and Ebeling, in preparation), but for this work we simply employ the
numerical probabilities plotted in the figure.

In addition to the extinction prior, we also require some assumptions about 
the probability of observing various galactic spectral types at various redshifts.
BPZ derives its default $P(z|T)$ prior from the Hubble Ultra-Deep Field;
however, since our area is so much larger than that of the HUDF, and since it -- by
design -- includes large-scale structure, one must adopt a broader prior that considers
galaxies over a wider range of redshifts and masses than those encountered in the HUDF.
In formulating an alternative to the default prior,
we must balance the need to rule out inherently improbable results
(such as observing an elliptical galaxy at $z=8$) with our limited
knowledge of galaxy evolution.

We first rule out high-redshift ellipticals. We 
construct a Schechter luminosity function for elliptical galaxies using the following parameters,
derived by \citet{Nakamura2003} from the Sloan Digital Sky Survey (SDSS): $M^*(r^*) = -21.52$;
$\alpha = -0.83$; and $\phi^* = 1.61 \times 10^{-3}$ Mpc$^{-3}$. We then take the BPZ elliptical
galaxy template, redshift it, account for intergalactic attenuation \citep{Madau1995}, and convolve it
with the F140W filter profile. Thus we derive an apparent magnitude for each combination of absolute
magnitude and redshift. By combining this information with our F140W limiting magnitude and the
elliptical galaxy luminosity function, we obtain a prior for observing elliptical galaxies which vanishes
smoothly around $z=3.5$.

This constraint on elliptical galaxy visibility follows directly from the weakness of their UV emission
combined with their empirically determined maximum luminosities.
In recognition of the fact that the luminosity functions evolve with cosmic time (e.g.,
\citealp{Bouwens2011,Bouwens2012}), we assume flat priors for the other galactic spectral types.
In particular, we have modified BPZ so that for each redshift, it reports the goodness-of-fit
probability for the \emph{most likely} galaxy template only, rather than summing the probabilities over all
templates. (See also the discussion in Section~\ref{sec:BPZhypz} concerning comparison of BPZ results with results from
hyperz.) Thus we consider only how closely the observed SED fits a galaxy template at high redshift without
taking into account the (unknown) likelihood of each template.

To validate the utility of this modification, we next determined which procedure (original or modified BPZ)
best reproduces the results of CLASH using only our four passbands. Since CLASH utilizes 16 passbands
for their photometric redshift determinations, their multiple bands in essence function as
very low-resolution spectroscopy. Thus, reproducing their redshifts would enhance our confidence
that our procedure yields reliable results.

We began with 47 high-redshift galaxies from tabs.~5 and 6 of
\citet{Bradley2013}. For each galaxy, \citeauthor{Bradley2013} report the photometric redshift
estimate along with the 95 per cent ($2\sigma$) confidence interval. We obtained the
magnitudes of each galaxy in our four passbands from the CLASH source
catalog\footnote{\url{https://archive.stsci.edu/prepds/clash/}}. We then ran both the
original implementation of BPZ and our modified version (using the aforementioned
prior on elliptical galaxies in both cases) on these 47 objects using only the four passbands considered
in this project. From the BPZ results, we determined the most likely redshifts and 68 per cent
($1\sigma$) confidence intervals. Since we only use four passbands whereas CLASH uses 16,
we expect their $2\sigma$ confidence intervals to be roughly comparable to our
$1\sigma$ intervals. Our limited number of passbands means that, in many cases,
95 per cent confidence intervals derived from our work would be so broad as to be almost useless.
Thus for our own results we quote 68 per cent intervals and, as noted in Section~\ref{sec:concl},
recommend spectroscopic follow-up for our most plausible candidates.

\begin{figure}
    \leavevmode\epsfxsize=8.5cm\epsfbox{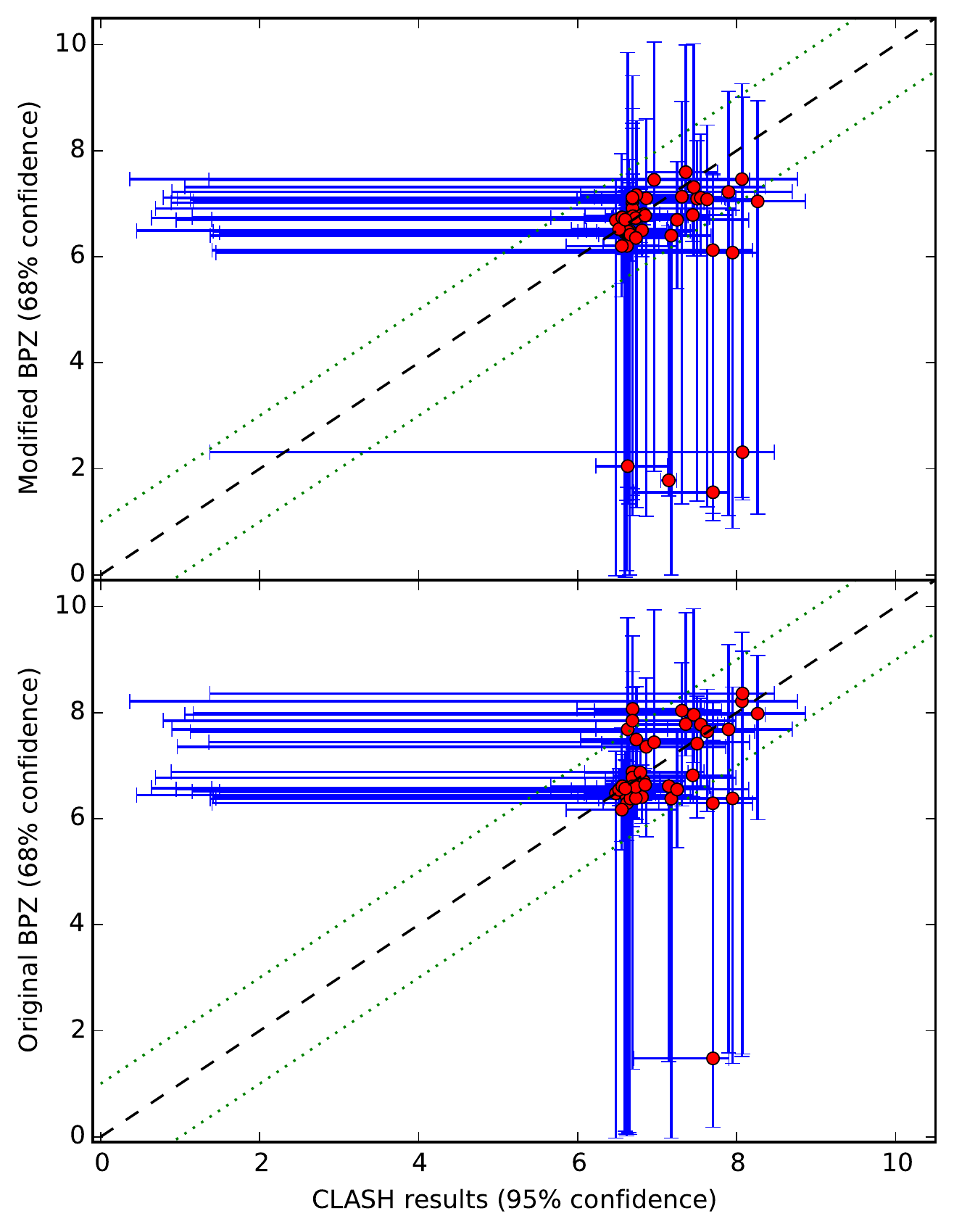}
    \caption{Comparison of photometric redshifts derived from CLASH with those derived from four passbands with BPZ (both the
    original version and as modified -- see Section~\ref{sec:SEDfit}). Green dotted lines
    are $\Delta z=1$ away from the main diagonal $z_\mathrm{4bands}=z_\mathrm{CLASH}$.
    The results of the original BPZ are more symmetric about the diagonal---and have fewer catastrophic
    outliers. However, the modified version produces more conservative results, in that it is less likely to
    produce photometric redshifts significantly in excess of the (presumably more nearly correct) CLASH estimates.}
\label{fig:BPZ_CLASH}
\end{figure}

The results appear in Fig.~\ref{fig:BPZ_CLASH}. We see that both versions of BPZ, when operating
on only four passbands, give highest-likelihood redshifts roughly comparable to those obtained by
CLASH. The original version of BPZ, as might be expected, produces a more symmetric scatter about
the diagonal $z_\mathrm{4bands} = z_\mathrm{CLASH}$; it also results in fewer `catastropic
outliers.' However, the results of the modified version are more conservative in that they seldom produce
redshifts significantly in excess of the CLASH results. The greatest excess redshift (compared with CLASH)
is $\Delta z=0.5$ for the modified version, as opposed to $\Delta z = 1.5$ for the original
version.

In two other respects our approach is conservative: first, we apply our extinction prior
to all spectral types of galaxies, thus assuming that dusty ellipticals are as likely as dusty starbursts
or spirals. Second, we use the SDSS $z=0$ luminosity function for ellipticals to estimate the likelihood
of observing such galaxies at higher redshifts, thus neglecting evolution. As a result of these two assumptions,
our estimated likelihood for dusty $z \sim 2$ ellipticals is probably higher than that found in the actual universe.
Given our limited number of passbands, our primary concern is to obtain conservative
photometric redshifts; we are thus willing to accept results which might underestimate the true redshift.
Hence we use the modified version of BPZ, with the understanding that we probably underestimate to
some degree the number of galaxies in each high-redshift bin.

\subsection{Eliminating Stars and Sub-stellar Objects}
\label{sec:spectest}
Section~\ref{sec:analysis} outlines our two-fold approach to eliminating stars and sub-stellar objects
from our list of dropouts. The first aspect of our approach is the morphological criterion displayed in Fig.~\ref{fig:stargal};
any objects in the light green portions of this figure we accept as galaxies. However, objects in the blue portions of the
figure have ambiguous morphology, and for these objects we use BPZ to determine how well their photometry
matches what one would expect for M-stars and brown dwarfs.

To do so, we obtained composite spectra of M-, L-, and T-dwarfs by stacking 25 spectra \citep{Burgasser2004,
Burgasser2006, Burgasser2007b, Burgasser2007a, Burgasser2008, Burgasser2010, Chiu2006, Kirkpatrick2010} of these
objects obtained from the SpeX Prism Spectral Libraries.\footnote{\url{http://pono.ucsd.edu/~adam/browndwarfs/spexprism/
library.html} We then included these spectra in the BPZ template list and, for these three spectral templates only,
imposed a delta-function prior limiting them to $z = 0$.}

We then prepared simulated stellar observations by convolving the model spectra with the HST filter profiles,
scaling to magnitudes typical of our candidates and adding uncertainties typical of our candidates. Experimentation
showed that, when applied to these simulated stars, BPZ with the extra templates yielded a probability spike
at $z = 0$ that was typically at least 30 per cent of the maximum height of the distribution.

 Thus, to impose our spectral criterion, we ran BPZ (with the extra templates)
on each of our ambiguous candidates and eliminated those for which the
probability density at $z = 0$ was at least 30 per cent of the maximum probability density. The remaining
objects we retained as galaxies. Finally, we removed the stellar templates from BPZ's template library and re-ran
it on these retained objects to obtain the probabilities reported in Table~\ref{tab:galprobs} and shown in
Figs~\ref{fig:galinfo1} and \ref{fig:galinfo2}.

\section{Discussion}

\subsection{SED-fit Codes}
\label{sec:BPZhypz}
There seems to be no `standard' code for SED-fitting. \citet{Bradley2013} use
BPZ, whereas \citet{Zheng2014} use a combination of BPZ and \texttt{iSEDfit}.
\citet{Atek2014} use \texttt{hyperz};
\citet{Bowler2014} use \texttt{LePhare}; and \citet{Oesch2013} use \texttt{ZEBRA} and \texttt{EAZY}.
Others \citep{McLure2011, McLure2013, Ellis2013} use proprietary code.

Since there is no accepted best SED-fit code (see \citealt{Hildebrandt2010} for a review and evaluation of 17 photometric
redshift codes) we thought it useful to compare the results of the
two codes with which we are most familiar, namely BPZ (modified as described above) and
hyperz\footnote{\url{http://webast.ast.obs-mip.fr/hyperz/}} \citep{Bolzonella2000}. BPZ marginalizes
over a carefully selected set of galaxy templates (a feature which our modification largely circumvents)
and naturally accommodates a prior distribution for those templates. It does not naturally handle various
levels of intrinsic extinction, requiring us to handle extinction as described in Section~\ref{sec:SEDfit}.
On the other hand, hyperz
fits not only redshift but also extinction and metallicity; however, it does not seem to accommodate
a prior on these variables.

\begin{figure}
    \leavevmode\epsfxsize=8.5cm\epsfbox{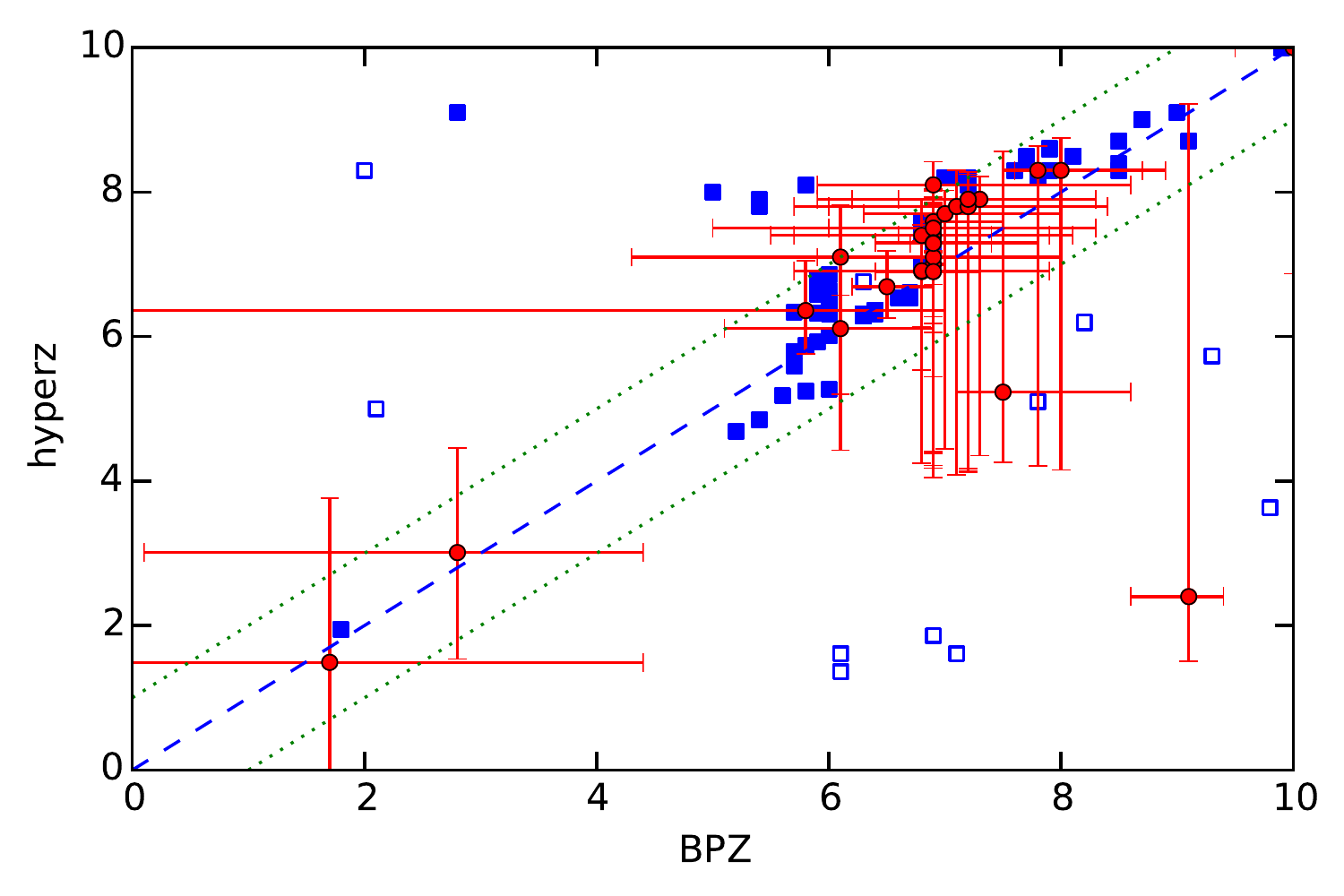}
    \caption{Photometric redshifts produced by BPZ (modified as
    described in Section~\ref{sec:SEDfit}) and by hyperz. Blue squares indicate probability distributions
    from BPZ which are bimodal at the 68 per cent level (i.e., the 68 per cent confidence regions are
    topologically disconnected), with only the most likely redshift plotted for each candidate. Red
    circles indicate unimodal distributions (connected 68 per cent confidence intervals). Filled markers
    indicate BPZ results which are consistent (to 68 per cent) with hyperz results; open markers indicate
    inconsistent results. Dotted lines show a distance of $\Delta z=1$ from the main diagonal. For readability,
    we show error bars for consistent and unimodal distributions only. We use only the red data points
    (redshift probabilities unimodal at 68 per cent) for the remainder of our analysis.
    }
\label{fig:BPZ_hyperz}
\end{figure}

Thus we compare the results of two approaches: the first is a modified BPZ with an extinction prior
and a limited galaxy template prior to disallow high-redshift ellipticals; the second is (unmodified) hyperz. The output
appears in Fig.~\ref{fig:BPZ_hyperz}. We note first the high incidence of objects (blue points) to which
the codes assigns a probability distribution which is bimodal at a 68 per cent level (meaning that the 68 per cent
confidence region is a topologically disconnected set). The majority of objects for which the 68 per cent 
BPZ and hyperz results differ (open circles) are these bimodal objects with poorly constrained redshifts.
Second, as in the comparison with the
CLASH results, we see that our modified version of BPZ is conservative in that it tends (with a few exceptions)
to assign a \emph{lower} redshift to (unimodal) objects than does hyperz. Thus, consistent with our
conservative approach, we shall henceforth ignore all objects to which BPZ assigns a bimodal distribution
at the 68 per cent level -- i.e., we shall consider only the red points plotted in Fig.~\ref{fig:BPZ_hyperz}.

\subsection{Magnification}

Determining the rest-frame UV magnitude of dropouts requires knowledge of
the degree to which gravitational lensing has magnified the object in question. Determination of the
lensing magnification requires in turn knowledge
of the mass distribution of the foreground cluster.

Mass maps based on spectroscopy of strong-lensing features
are available for a fraction of our target clusters, i.e. those for which such
features have been spectroscopically confirmed. 
Magnification  estimates  were derived by means of \texttt{Lenstool} \citep{Lenstool}
from the positions of strong-lensing features and a parametric model 
which includes the contribution from cluster-scale and galaxy-scale 
haloes (following the same approach as, e.g., \citealp{Richard2010, 
Limousin2015}). The available models allowed us to calculate the
magnification due to cluster gravity for the ten high-redshift
candidates presented in Table~\ref{tab:mags}. Note that the magnification errors include 
the statistical error from the parametric model but no
 systematic uncertainties from different modelling approaches. Although the uncertainties listed in Table~\ref{tab:mags} also include errors propagated from redshift uncertainties, they should thus be regarded as lower limits.

Some of these objects appear to be extremely luminous. Magnitudes below $-22$ are not unheard of;\footnote{An example would be the object at $\alpha = 325.0753169^\circ$, $\delta=-23.6772512^\circ$ in table~4 of \citet{Bradley2013}, which yields a restframe $M_{1500}$
around $-23$, although the authors note a small possibility that the unresolved source might be a star.}
however, \cite{Bouwens2011} note that UV magnitudes below $-24$ are physically
unlikely, given that the vigorous star formation required for such luminosity would quickly
fill the galaxy with dust. One of our objects
(MACSJ2135-0763) approaches this limit of $-24$. One possible explanation is galaxy-galaxy lensing, given
the nearby lower-redshift galaxy visible immediately to the left (in the third row of Fig.~\ref{fig:galinfo2}). Another possibility is
that the photometric redshift, despite being constrained by BPZ at the 68 per cent level, is in error. Inspection of its
redshift probability distribution (Fig.~\ref{fig:galinfo2}) shows a non-negligible probability of $z \sim 2$; nevertheless, the
probability of $z \ge 6.5$ is around 85 per cent (see Table~\ref{tab:galprobs}). This object would merit additional investigation.

\begin{table}
  \caption{Magnifications obtained for high-redshift candidates}
  \label{tab:mags}
  \begin{minipage}{8.5cm}
  \begin{center}
    \leavevmode
    \begin{tabular}{cccc} \hline
Object							& Photo-				& Ampli-				& Nominal  \\
ID\footnote{See Table~\ref{tab:galprobs}.}	& $z$\footnote{68 per cent confidence intervals.}				&  fication\footnote{Derived
from \texttt{Lenstool} \citep{Lenstool} models.}& $M_{1500}$\footnote{Uncertainties propagated from photo-$z$ uncertainties.} \\ \hline
MACSJ0140-0851  & $7.6^{+1.0}_{-0.2}$ & $1.26\pm0.00$  &  $-22.8\pm0.3$ \\
MACSJ0140-1028  & $6.1^{+1.4}_{-1.5}$ & $7.04\pm1.46$  &  $-18.4\pm0.7$ \\
MACSJ0152-0651  & $6.7^{+0.4}_{-0.4}$ & $10.03\pm0.26$  &  $-18.6\pm0.2$ \\
MACSJ0152-0871  & $6.9^{+0.1}_{-0.1}$ & $7.25\pm0.03$  &  $-21.9\pm0.0$ \\
MACSJ0712-0608  & $7.0^{+0.8}_{-0.8}$ & $1.24\pm0.00$  &  $-21.1\pm0.2$ \\
MACSJ0712-0699  & $8.0^{+1.9}_{-0.5}$ & $1.51\pm0.01$  &  $-23.1\pm0.5$ \\
MACSJ0947-0072  & $6.5^{+0.4}_{-0.3}$ & $1.17\pm0.00$  &  $-22.2\pm0.1$ \\
MACSJ1133-0922  & $7.2^{+1.1}_{-1.0}$ & $1.28\pm0.00$  &  $-21.3\pm0.2$ \\
MACSJ2135-0509  & $6.9^{+0.7}_{-0.4}$ & $5.15\pm0.10$  &  $-19.7\pm0.2$ \\
MACSJ2135-0763  & $9.1^{+0.3}_{-0.5}$ & $1.41\pm0.00$  &  $-23.8\pm0.1$ \\

 \hline
 \end{tabular}
 \end{center}
\end{minipage}
\end{table}

\begin{figure*}
   \leavevmode\epsfxsize=18cm\epsfbox{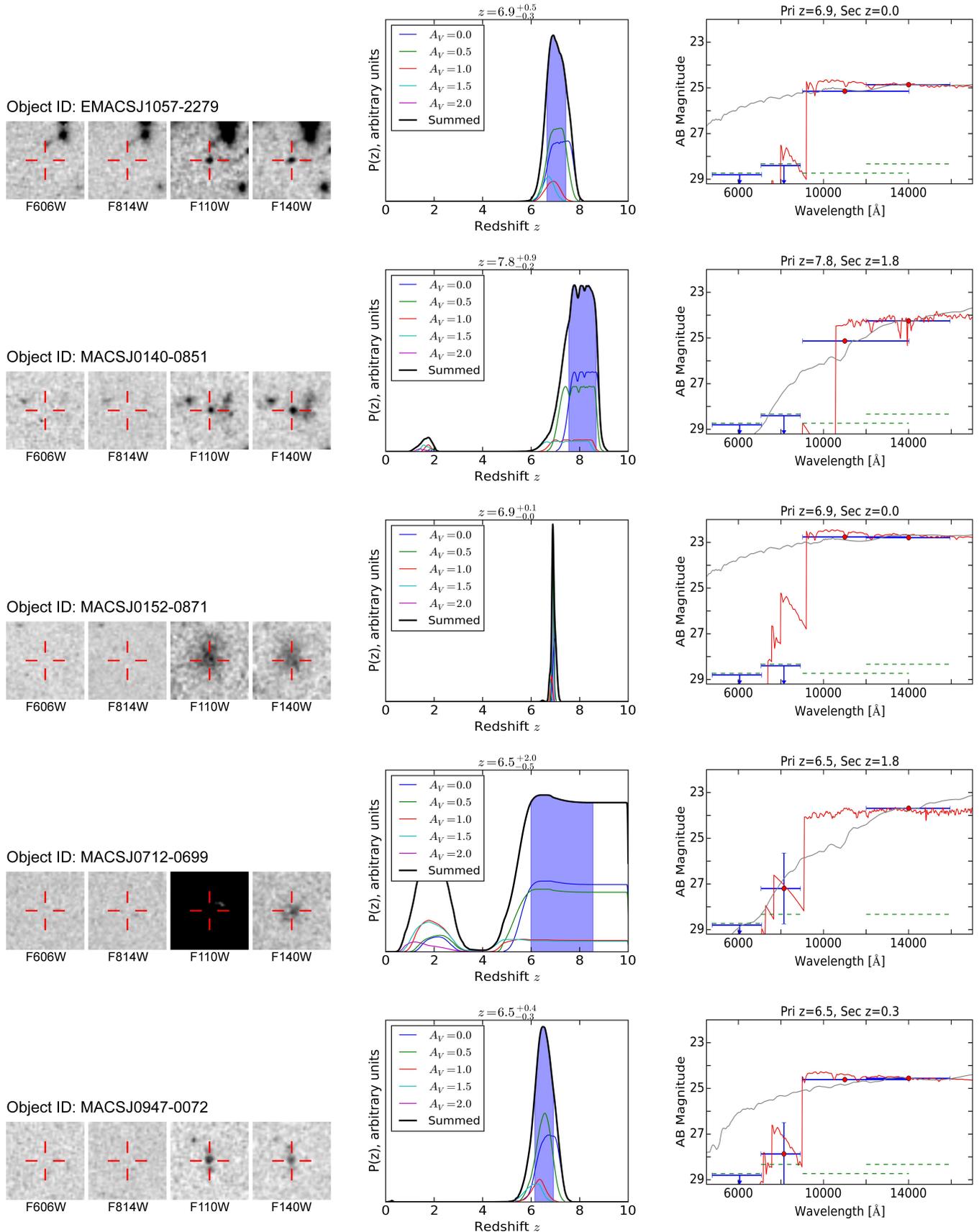}
    \caption{Images (5 arcsec per side), redshift probability distributions, and best-fitting
SEDs for selected dropout galaxies. For object IDs see Table~\ref{tab:galprobs}.
Blue shading denotes 68 per cent confidence (also noted above the
probability plots). Red SEDs show the most probable fit; grey SEDs show
the most probable low-redshift fit. The best-fitting template for each case
appears above the SED plot. Green dotted lines show limiting magnitude
in each filter.}
\label{fig:galinfo1}
\end{figure*}

\begin{figure*}
   \leavevmode\epsfxsize=18cm\epsfbox{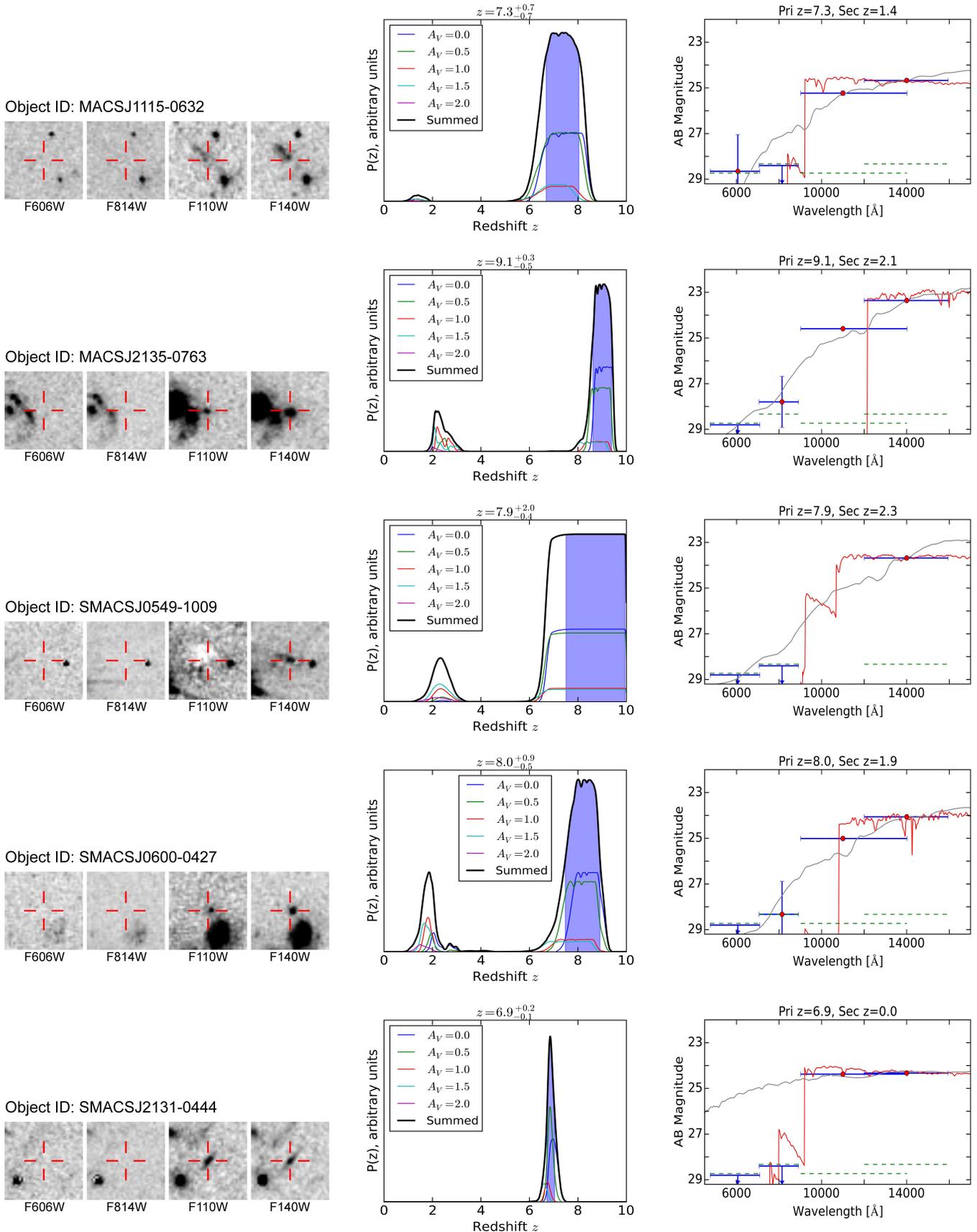}
    \caption{Images (5 arcsec per side), redshift probability distributions, and best-fitting
SEDs for selected dropout galaxies. For object IDs see Table~\ref{tab:galprobs}.
Blue shading denotes 68 per cent confidence (also noted above the
probability plots). Red SEDs show the most probable fit; grey SEDs show
the most probable low-redshift fit. The best-fitting template for each case
appears above the SED plot. Green dotted lines show limiting magnitude
in each filter.}
\label{fig:galinfo2}
\end{figure*}

\section{Results}
\label{sec:results}
Given our extinction prior, our galaxy type prior, and our modified version of BPZ, we determine both
a posterior probability distribution and a best-fitting SED for each of the dropout
galaxies. We retain only the 50 objects which meet the following criteria. First, the BPZ-derived probability
distribution must peak at $z > 5.5$. Second, the probability distribution must be unimodal at the 68 per cent level;
by this we mean that the 68 per cent confidence region (shaded blue in Figs.~\ref{fig:galinfo1} and \ref{fig:galinfo2}) must be
a connected set, so that only one peak rises to 68 per cent significance. Table~\ref{tab:galprobs} lists all 50
of these $I_{814}$-dropout galaxies, their magnitudes (corrected for Milky Way extinction) in each of the four
passbands; their photometric redshifts; and their probabilities of falling into redshift bins $z \sim 7$,
$z \sim 8$, and $z \sim 9$.

Note that two of these objects (EMACSJ1057-2279 and EMACSJ1057-2476 in Table~\ref{tab:galprobs}) appear to be
multiply lensed images of the same source. Both of these objects appear in the top row of Fig.~\ref{fig:galinfo1};
the first is in the centre of the stamp; the other is barely visible at the right-hand edge of the stamp. We make no attempt to determine the number of other objects in our
sample which might be multiply lensed, and in this work we do not account for this source of systematic
error in our sums of objects in each redshift bin.

The F110W photometry of four of these objects is defective, as noted in Section~\ref{sec:analysis}. The objects
appear at the end of Table~\ref{tab:galprobs}; in addition, one of them appears in the fourth row of
Fig.~\ref{fig:galinfo1}, and another appears in the fourth row of Fig.~\ref{fig:galinfo2}.
For these
objects, BPZ yields a relatively flat probability distribution at higher redshifts; in such cases, we weight
the probability of galaxies' placement into redshift bins with the summed probabilities for the other
dropout galaxies -- in essence using the other galaxies' redshifts as a prior for those with defective F110W photometry.
Figs.~\ref{fig:galinfo1} and \ref{fig:galinfo2} show the images, probability distributions, and SEDs
for some of the galaxies with the greatest likelihood of lying at a high redshift.

By summing the probabilities in Table~\ref{tab:galprobs}, we obtain the total
number of galaxies detected in each bin (also reported in Table~\ref{tab:galprobs}),
which round to 19 galaxies at $z \sim 7$, 6 at $z \sim 8$, and 3 at $z \sim 9$.
We thus estimate a total number of 27 high-redshift galaxies, which is one less than the
sum of the above numbers due to rounding. 
(See the table for 68 per cent Poisson confidence intervals.)
These sums include one galaxy (MACSJ1319-0075) for which the BPZ confidence interval $z = 9.3^{+0.5}_{-0.7}$
is inconsistent with the hyperz confidence interval $z = 5.7^{+0.6}_{-3.5}$. This object corresponds to the
single open red circle in Fig.~\ref{fig:BPZ_hyperz} and appears with a footnote in the table.

\begin{figure}
    \leavevmode\epsfxsize=8.5cm\epsfbox{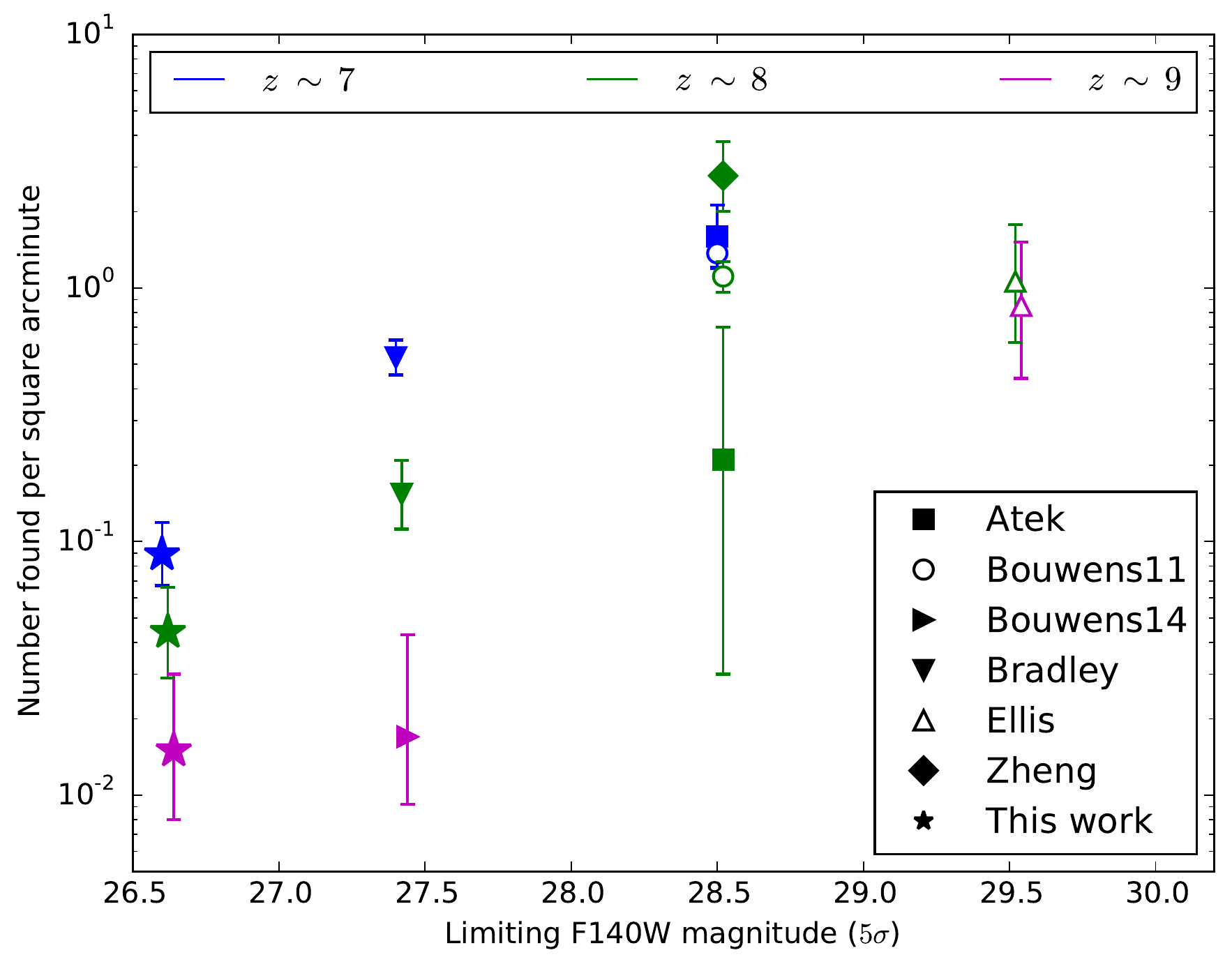}
    \caption{Comparison of the number of objects found in this work to that detected by
other surveys, as a function of limiting magnitude. Filled symbols denote lensed surveys,
and unfilled symbols denote unlensed surveys. (See Table~\ref{tab:othersresults}
for details.) Blue markers indicate objects
at $z \sim 7$; green, at $z \sim 8$; and magenta, at $z \sim 9$. Markers for $z \sim 8$
and $z \sim 9$ are slightly offset horizontally for clarity. Error bars
represent Poisson confidence intervals. Note that for lensed surveys the vertical axis lists the number of objects
per unit solid angle in the image plane, not the source plane.}
\label{fig:othersresults}
\end{figure}

Finally, in order to evaluate the effectiveness of our approach, we compare the number of high-redshift
galaxies estimated in this survey with those detected by others.
One would expect a positive correlation between the number of galaxies and the
limiting magnitude of the survey. Using a thousand 0.4$''$-diameter apertures in each of our
clusters, we determine that the $5\sigma$ limiting magnitude for our survey in F140W is 26.6.
Normalizing the number of detected high-redshift objects to the solid angle coverage of 
various searches, and plotting the results against limiting magnitudes, we obtain Fig.~\ref{fig:othersresults}.
(See also Table~\ref{tab:othersresults} for the fields involved.)

Note that this figure combines surveys with disparate parameters (e.g., various detection bands, differing
uniformity of depth, lensed/unlensed, etc.); thus one ought not to attempt to derive any sort of cumulative luminosity function from it.
However, what Fig.~\ref{fig:othersresults} does show is that our survey returns a number of galaxies commensurate with our low limiting magnitude. It is the large solid angle of our survey that permits discovery of a non-negligible number of high-redshift candidates, underlining that shallow, wide surveys efficiently probe the bright end of the high-redshift UV luminosity function.

\begin{table}
  \begin{minipage}{8.5cm}
  \centering
  \caption{Surveys plotted in Fig.~\ref{fig:othersresults}}
  \label{tab:othersresults}
  \begin{tabular}{llcc} \hline
				& Survey/			&Depth$^a$		& Solid Angle\\
Reference			&	Field(s)			&($5\sigma$)		& (arcmin$^2$)\\\hline
\citet{Atek2014}		&	Abell-2744 (HFF)	&28.5			& 4.7\\	      
\citet{Bouwens2011}	&	HUDF09, ERS		&28.5			& 53\\	
\citet{Bouwens2014}	&	CLASH			&27.4			& 77\\	
\citet{Bradley2013}	&	CLASH			&27.4			& 82\\	
\citet{Ellis2013}		&	HUDF				&29.5			& 4.7\\	
\citet{Zheng2014}	&	Abell-2744 (HFF)	&28.5			& 4.7\\	
This work			&	MACS			&26.6			& 135\\
 \hline
\end{tabular}
\end{minipage}
$^a$Limiting magnitude in F140W filter, or interpolated between F125W and F160W limiting magnitudes if
F140W not utilized. If depths differ across fields in the same survey, we cite an average depth weighted
by the solid angle of the fields involved.\\
$^b$In image plane, not source plane.
\end{table}

\section{Conclusions}
\label{sec:concl}
We draw the following conclusions from our survey. First, broad and relatively
shallow surveys like ours can effectively detect high-redshift galaxy candidates. Devoting only
about an hour of Hubble time to each of 29 clusters, we were able to identify 50
objects which have a probability distribution which peaks at high redshift and is
unimodal at the 68 per cent level. Of these objects, we estimate that 27 
lie at redshift $z \ga 7$.
Since our survey is relatively shallow, we are therefore sampling the bright end of the UV galaxy
luminosity function at such redshifts. Our currently available mass maps suggest
the existence of several extremely luminous objects with a photometric redshift of
$z \sim 8$--9, placing them only about 600 Myr after decoupling. Such objects are ideal
targets for further study, if for no other reason than to confirm or falsify
the high photometric redshifts.

This further investigation is important for at least three reasons. First, despite our rejection
of objects with significantly bimodal distributions, inspection of Table~\ref{tab:galprobs} shows that
almost all of the remaining objects still have non-negligible lower-redshift probabilities. Indeed,
the summed likelihoods show that we can
expect only 27 of the 50 listed objects to actually lie at $z \ga 7$. Second, we derived our
extinction prior from Lyman-break galaxy data, whereas low-redshift F814W dropouts represent a different
population (4000-\AA\ break galaxies and possibly some high-luminosity infrared galaxies). This population
mismatch at lower redshifts could potentially introduce a systematic bias into our results. Third,
our analysis of BPZ and hyperz shows that the two codes can occasionally yield wildly divergent results. Thus
the choice of a specific SED-fit routine can introduce additional -- and probably significant --
systematic bias. We are unable to quantify these biases, and the resulting intrinsic
uncertainty is not captured in the statistical error bars we report.

We note, however, that the low-redshift interlopers are scientifically interesting in their own right. Over 70 of our 124 F814W-dropout galaxies
do not appear in Table~\ref{tab:galprobs} due to bimodalities in their probability distributions;
thus, besides high-redshift candidates like those we report in this work, surveys such as ours can
identify a large number of lower-redshift objects conducive to the study of galaxy evolution.

Finally, we conclude that even as few as four broad-band filters suffice to isolate
promising objects. However, our results also reinforce
the importance of SED fitting in obtaining reliable photometric redshifts: dependence on
a simple dropout criterion alone would have doubled our catalog size by introducing
objects with a significant likelihood of lying at low redshift. SED-fitting is especially important
for surveys which utilize a small number of passbands; a greater number of
passbands allows the use of more complex colour criteria (e.g., \citealt{Castellano2010a}) which
in practice function as a coarse SED fit. In particular, the F606W band was necessary to
eliminate high equivalent-width emission-line galaxies from our sample. The small number
of passbands did lead to a large number of objects with significantly bimodal photometric
redshifts, which we eliminated from consideration; it was the large solid angle of the
survey which permitted isolation of a significant number of more securely identified
objects.

Spectroscopic investigations of some of our best candidates promise to better constrain
these objects' redshifts, as do deeper space-based observations with more filters, and/or slitless near infrared spectroscopy with WFC3 grisms. In addition, the eMACS SNAPs program proposes to survey 50 massive clusters with $z > 0.5$.
Application of our method to this and other future programs has the potential to isolate yet more promising candidates.
Thus we confidently expect that the study of high-redshift
galaxies will continue to shed light on the processes occurring in the early universe.

\section*{Acknowledgement}
We thank Anton Koekemoer for his invaluable advice concerning DrizzlePac parameters, weight
maps, and photometric errors.
We also thank Hakim Atek for helpful advice regarding galaxy templates and extinction corrections. 
AR and HE gratefully acknowledge financial support from STScI grants GO-10491, GO-10875,
GO-12166, and GO-12884.

\bibliographystyle{mn2e}
\bibliography{MACSHighRedshift}

\begin{table*}
 \caption{Dropout galaxies; all confidence intervals are 68 per cent.}
 \label{tab:galprobs}
 \begin{minipage}{18.5cm}
 \begin{center}
    \leavevmode
    \begin{tabular}{lcccccccccc} \hline
ID	& 	$\alpha_\mathrm{J2000}$	&	$\delta_\mathrm{J2000}$		&	$m_\mathrm{F606W}$	&	$m_\mathrm{F814W}$	&	$m_\mathrm{F110W}$	&	$m_\mathrm{F140W}$	&	$z$\footnote{68 per cent confidence intervals.}		&	$P_7$\footnote{$P_7$, $P_8$, and $P_9$ denote the probabilities that the object falls within redshift bins $[6.5,7.5)$, $[7.5, 8.5)$, or $[8.5, 9.5)$, respectively.}	&	$P_8$$^{\textstyle b}$	&	$P_9$$^{\textstyle b}$ \\
\hline
\multicolumn{11}{l}{\textbf{Dropouts with reliable photometry in all four bands}\rule[-2ex]{0cm}{5ex}}\\
EMACSJ1057-2261  & $10^{\mathrm{h}}57^{\mathrm{m}}32.02^{\mathrm{s}}$  & $57^\circ59'30''$  & $> 28.8$  & $27.48 \pm 0.69$  & $26.04 \pm 0.13$  & $26.17 \pm 0.20$  & $6.1^{+0.8}_{-1.0}$    & 0.19   & 0.01   & 0.00  \rule[-2ex]{0cm}{2ex}\\
EMACSJ1057-2279\footnote{Part of multiply-lensed system; see Section~\ref{sec:results}.}  & $10^{\mathrm{h}}57^{\mathrm{m}}27.80^{\mathrm{s}}$  & $57^\circ59'7''$  & $> 28.8$  & $> 28.4$  & $25.15 \pm 0.06$  & $24.86 \pm 0.06$  & $6.9^{+0.5}_{-0.3}$    & 0.79   & 0.14   & 0.00  \rule[-2ex]{0cm}{2ex}\\
EMACSJ1057-2476$^{\textstyle c}$  & $10^{\mathrm{h}}57^{\mathrm{m}}27.48^{\mathrm{s}}$  & $57^\circ59'7''$  & $> 28.8$  & $> 28.4$  & $24.55 \pm 0.03$  & $24.15 \pm 0.03$  & $6.9^{+0.6}_{-0.1}$    & 0.74   & 0.25   & 0.00  \rule[-2ex]{0cm}{2ex}\\
MACSJ0140-0851  & $1^{\mathrm{h}}39^{\mathrm{m}}58.63^{\mathrm{s}}$  & $-5^\circ56'12''$  & $> 28.8$  & $> 28.4$  & $25.13 \pm 0.05$  & $24.25 \pm 0.03$  & $7.8^{+0.9}_{-0.2}$    & 0.19   & 0.60   & 0.17  \rule[-2ex]{0cm}{2ex}\\
MACSJ0140-1028  & $1^{\mathrm{h}}40^{\mathrm{m}}3.70^{\mathrm{s}}$  & $-5^\circ55'15''$  & $> 28.8$  & $28.04 \pm 0.97$  & $26.38 \pm 0.19$  & $26.25 \pm 0.24$  & $6.1^{+1.7}_{-1.8}$    & 0.19   & 0.09   & 0.01  \rule[-2ex]{0cm}{2ex}\\
MACSJ0152-0477  & $1^{\mathrm{h}}52^{\mathrm{m}}33.16^{\mathrm{s}}$  & $-28^\circ54'42''$  & $> 28.8$  & $> 28.4$  & $26.62 \pm 0.16$  & $26.35 \pm 0.17$  & $6.9^{+1.2}_{-1.4}$    & 0.32   & 0.17   & 0.01  \rule[-2ex]{0cm}{2ex}\\
MACSJ0152-0651  & $1^{\mathrm{h}}52^{\mathrm{m}}34.36^{\mathrm{s}}$  & $-28^\circ54'27''$  & $> 28.8$  & $> 28.4$  & $25.89 \pm 0.11$  & $25.88 \pm 0.16$  & $6.8^{+0.5}_{-0.4}$    & 0.67   & 0.07   & 0.00  \rule[-2ex]{0cm}{2ex}\\
MACSJ0152-0871  & $1^{\mathrm{h}}52^{\mathrm{m}}36.05^{\mathrm{s}}$  & $-28^\circ54'35''$  & $> 28.8$  & $> 28.4$  & $22.76 \pm 0.02$  & $22.79 \pm 0.03$  & $6.9^{+0.1}_{-0.0}$    & 1.00   & 0.00   & 0.00  \rule[-2ex]{0cm}{2ex}\\
MACSJ0257-0913  & $2^{\mathrm{h}}57^{\mathrm{m}}41.18^{\mathrm{s}}$  & $-22^\circ10'10''$  & $> 28.8$  & $> 28.4$  & $26.23 \pm 0.15$  & $25.89 \pm 0.14$  & $6.9^{+1.1}_{-0.9}$    & 0.37   & 0.23   & 0.01  \rule[-2ex]{0cm}{2ex}\\
MACSJ0712-0608  & $7^{\mathrm{h}}12^{\mathrm{m}}13.17^{\mathrm{s}}$  & $59^\circ32'52''$  & $> 28.8$  & $> 28.4$  & $25.99 \pm 0.10$  & $25.58 \pm 0.10$  & $7.0^{+1.0}_{-0.7}$    & 0.43   & 0.25   & 0.00  \rule[-2ex]{0cm}{2ex}\\
MACSJ0947-0072  & $9^{\mathrm{h}}47^{\mathrm{m}}17.86^{\mathrm{s}}$  & $76^\circ24'21''$  & $> 28.8$  & $27.87 \pm 1.37$  & $24.62 \pm 0.05$  & $24.56 \pm 0.07$  & $6.5^{+0.4}_{-0.3}$    & 0.53   & 0.00   & 0.00  \rule[-2ex]{0cm}{2ex}\\
MACSJ1115-0329  & $11^{\mathrm{h}}15^{\mathrm{m}}15.94^{\mathrm{s}}$  & $53^\circ19'5''$  & $28.28 \pm 0.84$  & $> 28.4$  & $26.35 \pm 0.18$  & $26.34 \pm 0.25$  & $6.8^{+1.1}_{-1.1}$    & 0.39   & 0.13   & 0.01  \rule[-2ex]{0cm}{2ex}\\
MACSJ1115-0632  & $11^{\mathrm{h}}15^{\mathrm{m}}17.45^{\mathrm{s}}$  & $53^\circ20'27''$  & $28.65 \pm 1.60$  & $> 28.4$  & $25.23 \pm 0.08$  & $24.67 \pm 0.07$  & $7.3^{+0.7}_{-0.7}$    & 0.48   & 0.39   & 0.01  \rule[-2ex]{0cm}{2ex}\\
MACSJ1124-0811  & $11^{\mathrm{h}}24^{\mathrm{m}}28.77^{\mathrm{s}}$  & $43^\circ50'41''$  & $> 28.8$  & $> 28.4$  & $26.38 \pm 0.17$  & $25.89 \pm 0.15$  & $7.1^{+1.3}_{-1.4}$    & 0.28   & 0.25   & 0.04  \rule[-2ex]{0cm}{2ex}\\
MACSJ1133-0922  & $11^{\mathrm{h}}33^{\mathrm{m}}8.33^{\mathrm{s}}$  & $50^\circ8'27''$  & $28.63 \pm 1.39$  & $> 28.4$  & $25.98 \pm 0.13$  & $25.47 \pm 0.12$  & $7.2^{+1.1}_{-1.0}$    & 0.35   & 0.29   & 0.02  \rule[-2ex]{0cm}{2ex}\\
MACSJ1621-0860  & $16^{\mathrm{h}}21^{\mathrm{m}}23.01^{\mathrm{s}}$  & $38^\circ11'13''$  & $> 28.8$  & $> 28.4$  & $26.51 \pm 0.13$  & $26.22 \pm 0.14$  & $6.8^{+1.1}_{-1.1}$    & 0.36   & 0.17   & 0.00  \rule[-2ex]{0cm}{2ex}\\
MACSJ1652-0135  & $16^{\mathrm{h}}52^{\mathrm{m}}26.52^{\mathrm{s}}$  & $55^\circ34'43''$  & $> 28.8$  & $> 28.4$  & $26.81 \pm 0.19$  & $26.52 \pm 0.20$  & $6.9^{+1.4}_{-1.9}$    & 0.26   & 0.17   & 0.02  \rule[-2ex]{0cm}{2ex}\\
MACSJ2051-0806  & $20^{\mathrm{h}}51^{\mathrm{m}}13.80^{\mathrm{s}}$  & $2^\circ16'48''$  & $> 28.8$  & $> 28.4$  & $26.32 \pm 0.16$  & $26.11 \pm 0.18$  & $6.9^{+1.1}_{-1.0}$    & 0.39   & 0.18   & 0.01  \rule[-2ex]{0cm}{2ex}\\
MACSJ2135-0509  & $21^{\mathrm{h}}35^{\mathrm{m}}10.90^{\mathrm{s}}$  & $-1^\circ3'12''$  & $> 28.8$  & $> 28.4$  & $25.76 \pm 0.14$  & $25.52 \pm 0.16$  & $6.9^{+0.9}_{-0.5}$    & 0.53   & 0.23   & 0.00  \rule[-2ex]{0cm}{2ex}\\
MACSJ2135-0763  & $21^{\mathrm{h}}35^{\mathrm{m}}8.24^{\mathrm{s}}$  & $-1^\circ2'41''$  & $> 28.8$  & $27.80 \pm 1.11$  & $24.59 \pm 0.04$  & $23.36 \pm 0.02$  & $9.1^{+0.3}_{-0.5}$    & 0.00   & 0.07   & 0.78  \rule[-2ex]{0cm}{2ex}\\
MACSJ2135-1078  & $21^{\mathrm{h}}35^{\mathrm{m}}14.82^{\mathrm{s}}$  & $-1^\circ2'25''$  & $28.72 \pm 2.05$  & $> 28.4$  & $26.15 \pm 0.18$  & $25.69 \pm 0.16$  & $7.2^{+1.2}_{-1.2}$    & 0.31   & 0.26   & 0.05  \rule[-2ex]{0cm}{2ex}\\
SMACSJ0600-0180  & $6^{\mathrm{h}}0^{\mathrm{m}}9.21^{\mathrm{s}}$  & $-43^\circ53'6''$  & $> 28.8$  & $28.04 \pm 1.19$  & $25.48 \pm 0.11$  & $24.81 \pm 0.08$  & $6.9^{+1.7}_{-1.0}$    & 0.27   & 0.26   & 0.03  \rule[-2ex]{0cm}{2ex}\\
SMACSJ0600-0427  & $6^{\mathrm{h}}0^{\mathrm{m}}9.12^{\mathrm{s}}$  & $-43^\circ53'44''$  & $> 28.8$  & $28.33 \pm 1.44$  & $25.01 \pm 0.07$  & $24.06 \pm 0.04$  & $8.0^{+0.9}_{-0.5}$    & 0.11   & 0.48   & 0.25  \rule[-2ex]{0cm}{2ex}\\
SMACSJ2031-0768  & $20^{\mathrm{h}}31^{\mathrm{m}}45.98^{\mathrm{s}}$  & $-40^\circ37'31''$  & $> 28.8$  & $> 28.4$  & $26.38 \pm 0.13$  & $25.87 \pm 0.11$  & $7.2^{+1.1}_{-1.3}$    & 0.30   & 0.25   & 0.02  \rule[-2ex]{0cm}{2ex}\\
SMACSJ2131-0444  & $21^{\mathrm{h}}31^{\mathrm{m}}6.95^{\mathrm{s}}$  & $-40^\circ18'57''$  & $> 28.8$  & $> 28.4$  & $24.38 \pm 0.04$  & $24.33 \pm 0.06$  & $6.9^{+0.2}_{-0.1}$    & 0.98   & 0.00   & 0.00  \rule[-2ex]{0cm}{2ex}\\
SMACSJ2131-0516  & $21^{\mathrm{h}}31^{\mathrm{m}}6.47^{\mathrm{s}}$  & $-40^\circ18'51''$  & $> 28.8$  & $> 28.4$  & $25.07 \pm 0.08$  & $24.83 \pm 0.09$  & $6.9^{+0.5}_{-0.2}$    & 0.78   & 0.16   & 0.00  \rule[-2ex]{0cm}{2ex}\\

SMACSJ2131-0567  & $21^{\mathrm{h}}31^{\mathrm{m}}6.14^{\mathrm{s}}$  & $-40^\circ18'23''$  & $27.61 \pm 0.91$  & $> 28.4$  & $25.24 \pm 0.09$  & $24.45 \pm 0.06$  & $7.5^{+1.1}_{-0.4}$    & 0.35   & 0.46   & 0.11  \rule[-2ex]{0cm}{2ex}\\
\multicolumn{8}{l}{Summed probabilities for objects with reliable photometry in all four bands:} &   11.6	&	5.6	&	1.6	\rule[-2ex]{0cm}{0ex}\\

\multicolumn{11}{l}{\textbf{Dropouts with defective F110W photometry}\footnote{Reported probabilities for dropouts with defective F110W photometry are those
derived directly from the BPZ fit, before application of a prior based on the redshifts of the other dropouts (see text).}\rule[-2ex]{0cm}{5ex}}\\
MACSJ0712-0414  & $7^{\mathrm{h}}12^{\mathrm{m}}25.64^{\mathrm{s}}$  & $59^\circ31'55''$  & $> 28.8$  & $> 28.4$  & ---  & $22.72 \pm 0.02$  & $8.2^{+1.8}_{-0.4}$    & 0.21   & 0.31   & 0.31  \rule[-2ex]{0cm}{2ex}\\
MACSJ0712-0699  & $7^{\mathrm{h}}12^{\mathrm{m}}29.23^{\mathrm{s}}$  & $59^\circ32'59''$  & $> 28.8$  & $27.20 \pm 1.55$  & ---  & $23.69 \pm 0.05$  & $6.5^{+2.0}_{-0.5}$    & 0.18   & 0.17   & 0.17  \rule[-2ex]{0cm}{2ex}\\
SMACSJ0549-0900  & $5^{\mathrm{h}}49^{\mathrm{m}}16.25^{\mathrm{s}}$  & $-62^\circ5'15''$  & $> 28.8$  & $> 28.4$  & ---  & $24.35 \pm 0.05$  & $8.0^{+1.9}_{-0.9}$    & 0.23   & 0.24   & 0.24  \rule[-2ex]{0cm}{2ex}\\
SMACSJ0549-1009  & $5^{\mathrm{h}}49^{\mathrm{m}}14.87^{\mathrm{s}}$  & $-62^\circ5'46''$  & $> 28.8$  & $> 28.4$  & ---  & $23.68 \pm 0.04$  & $7.9^{+2.0}_{-0.4}$    & 0.23   & 0.28   & 0.28  \rule[-2ex]{0cm}{2ex}\\
SMACSJ0549-9147  & $5^{\mathrm{h}}49^{\mathrm{m}}11.29^{\mathrm{s}}$  & $-62^\circ5'7''$  & $28.56 \pm 2.22$  & $> 28.4$  & ---  & $24.62 \pm 0.07$  & $7.9^{+2.1}_{-0.9}$    & 0.22   & 0.23   & 0.23  \rule[-2ex]{0cm}{2ex}\\

\multicolumn{8}{l}{Summed probabilities for candidates with defective F110W photometry:\footnote{Summed probabilities reflect the prior described in the text
and referenced in the immediately preceding note.}} &0.7	&	0.4&	0.1\rule[-2ex]{0cm}{0ex}\\

\multicolumn{8}{l}{Net summed probability in each redshift bin:} & 		12.2 				&	5.9		&	1.7\rule[-2ex]{0cm}{0ex}\\
\multicolumn{8}{l}{\textbf{Net detections reported in each redshift bin:}}  &$\mathbf{12^{+4}_{-3}}$  	&$\mathbf{6^{+3}_{-2}}$	&$\mathbf{2^{+2}_{-1}}$\rule[-2ex]{0cm}{0ex}\\
\end{tabular}
\end{center}
\end{minipage}

\label{lastpage}
\end{table*}

\end{document}